\documentclass[aps,prd,reprint,floatfix,groupedaddress,nofootinbib]{revtex4-1}
\usepackage{graphicx,amsmath}
\begin{document}

\preprint{NSF-KITP-13-151}

\title{Dark Matter Annihilations in the Causal Diamond}


\author{Andrew Scacco}
%\thanks{}
\affiliation{University of California at Davis;
Department Of Physics\\
One Shields Avenue;
Davis, CA 95616}
\author{Andreas Albrecht}
%\thanks{}
\affiliation{University of California at Davis;
Department Of Physics\\
One Shields Avenue;
Davis, CA 95616}


\date{\today}

\begin{abstract}
%We investigate the implications of dark matter annihilations for
%cosmological parameter constraints using the
%causal entropic principle. In this approach cosmologies are
%weighted by the total entropy production within a
%causally connected region of spacetime. We calculate the expected
%entropy from dark matter annihilations within the causal diamond and investigate the preferred values of
%the cosmological constant and the mass of the annihilating dark matter
%and their dependence on the assumptions in the models. We typically
%find preferred values of $\Lambda $ on the order of $10^{-5}$ of the
%present value assuming dark matter annihilations are the primary
%source of entropy production. We also investigate the effect of
%combining this entropy with the entropy production from stars. The
%greatest amount of entropy production from dark matter within the
%causal diamond is likely to occur with light keV scale dark matter,
%thus favoring the keV dark matter mass scale.

%We also find that using the causal entropic principle on the entropy from stars by varying the cross section for dark matter annihilation produces better agreement with observations than varying either the baryon fraction or the total matter abundance independently, and also that for dark matter annihilation the increase in matter density in the early universe resulting from a lowered cross section causes the dark matter to annihilate earlier, favoring a larger value of cosmological constant.

We investigate the implications of dark matter annihilations for
cosmological parameter constraints using the
causal entropic principle. In this approach cosmologies are
weighted by the total entropy production within a
causally connected region of spacetime. We calculate the expected
entropy from dark matter annihilations within the causal diamond and investigate the preferred values of
the cosmological constant and the mass and annihilation cross section of the annihilating dark matter
and their dependence on the assumptions in the models. For realistic values of the cross section we typically
find preferred values of $\Lambda $ on the order of $10^{-5}$ of the
present value assuming dark matter annihilations are the primary
source of entropy production. The
greatest amount of entropy production from dark matter within the
causal diamond is likely to occur with light keV scale dark matter with low annihilation cross section. We also investigate the effect of
combining this entropy with the entropy production from stars, and show that if the primary source of entropy production is from stars, varying the dark matter cross section directly produces a preferred value of $\Omega_m$ in excellent agreement with observations.

%,
%which is just heavy enough to evade constraints from structure
%formation. 
\end{abstract}

\pacs{}
%\keywords{}

\maketitle

\tableofcontents

\section{Introduction}
\label{Sect:Intro}

A fundamental understanding of physical laws might yield a picture in
which the physical constants are fixed from first principles.
Alternatively some or all constants could take
on a range of values in some meta picture, making the values we
observe an ``environmental'' feature of physics.  This could be due to the existence of a
multiverse, in which our universe is but one of many universes that we
might be living in. In many of these multiverse theories, it is
possible (or at least hoped to be possible) to predict the distribution of
possible universes and their likelihoods. In this case, predictions
about values of constants in the universe should boil down to probability distributions, which we could compare
with observations using the usual statistical tools. 
A famous example of this later approach was used by Weinberg to
predict a small but positive value for cosmological
constant~\cite{weinberg1987} before something like that was observed.  This result (and
subsequent extensions motivated by the string theory
landscape~\cite{Bousso:2000xa}), has generated considerable interest
to this approach.  Such analysis usually requires some sort of
``anthropic'' reasoning which folds in some measure of the likelihood
of observers existing with different physical parameters. In general,
stating exactly what such conditions might be is tricky.  Bousso
et al.~\cite{bousso2007} proposed the causal entropic principle (CEP),
an elegant approach which weighs different parameter choices according to entropy
production in the corresponding cosmology.  Entropy increase
(the 2nd law) is a critical resource for any imaginable observer, even
ones extremely different from us.  So entropy production seems to have
something to do with counting observers, yet it also is a familiar
quantity that physicists are used to calculating in concrete terms. 

\begin{figure}
\includegraphics[width=\linewidth]{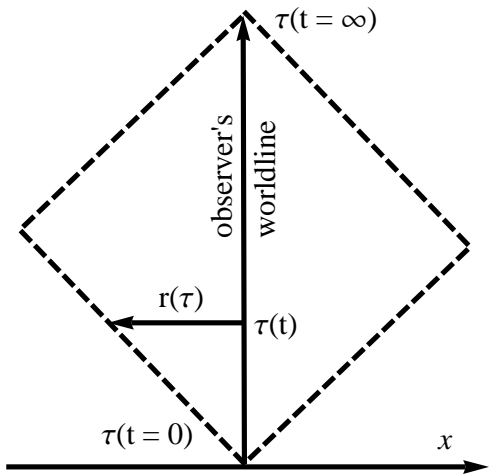}
\caption{The causal
diamond can be simply described as containing everything that an
observer can affect, which in turn can be later observed by that same
observer. So the boundaries are the future light cone of the beginning
of the observer's worldline and the past light cone of the end of the
observer's worldline. \label{fig:causal}} 
\end{figure}

These methods require entropy production to be measured in a
specific four-volume.  For the CEP, this volume
is the causal diamond, as shown in Fig.~\ref{fig:causal}.
%{\tt: Lets submit this to PRD,
  %which means referring to figures this way.  (Please change
  %throughout, except full word ``Figure'' is still used if at the
  %start of a sentence)}
The calculation of this volume in given in Sec.~\ref{sec:scale}.   

The causal entropic principle (CEP) has been used to find probability
distributions for the cosmological
constant, curvature, density contrast, baryon fraction, overall matter
abundance, and decays of dark
matter~\cite{cline2008,bozek2009}. Subtleties connected to inhomogeneities have also been
pointed out in~\cite{phillips2011}. Other work on the CEP can be found in~\cite{Bousso:2010im,Bousso:2010zi,Bousso:2010vi,Bousso:2009gx,Feldstein:2005bm,Garriga:2005ee,Maor:2008df,MersiniHoughton:2008ui,Salem:2009eh}.

The new work presented in this paper adds entropy production from dark
matter annihilations to the CEP calculations.  We calculate the joint
probability distribution for dark matter particle mass and $\Lambda$,
and show that (depending on parameters) dark matter annihilations can
significantly impact the results.  We discuss several interpretations,
including predicting a preferred value of the dark matter particle
mass of order $1$ keV if $\Lambda$ is fixed to the observed value.  
We find our work all the more timely due to recent hints that suggest the possibility
of detecting signatures from dark matter annihilations from gamma ray
lines, as in~\cite{bringmann2012}. 
There are tantalizing hints that
dark matter annihilations may have been seen by PAMELA~\cite{Adriani:2008zr}, ATIC~\cite{Chang:2008aa}, Fermi~\cite{Fermi-LAT:2013uma}, or AMS~\cite{Jin:2013nta}.
%\cite{abdo2009, Ackermann:2012qk, Adriani:2008zr, Jin:2013nta}.
It should be noted, however, that the typical dark matter particles
hypothesized to account for these lines are WIMPS at the GeV
scale. For simplicity we have chosen a specific family of dark matter
models for this work.  They are sterile neutrino models which as it
happens have been most systematically studied for keV scale dark
matter masses. We include extrapolations of these
models to GeV scales (which end up being disfavored).  Our 
extrapolated models are not the same as many models currently being
proposed in response to these experimental hints, but we expect their
general properties to be quite representative for our purposes.

The dark matter particle mass comes in in two important ways. Firstly,
in standard thermal production scenarios the dark matter mass is
related to the dark matter particle velocities and thus determines the
``streaming length'' on which cosmic structure is washed out by dark
matter motions.  Here we consider a range of parameters which run from
``cold dark matter'' for which the streaming length is so small as to
be unimportant for our considerations as well as ``warm dark matter'' where the
streaming affects some smaller scale aspects of cosmic structure (an
advantage, some argue, in providing consistency with observations~\cite{Weinberg:2013aya}). 
Additionally,
there are more particles per unit mass for lighter (warmer) dark
matter, meaning they will be more likely to find each other to
annihilate. Also, since warm dark matter disrupts formation of
structures on small scales, the balance of these effects allows for
the prediction of the dark matter mass. 

%{\tt
%Support for the presence of warm dark matter can be found in~\cite{deVega:2009ku,Song:2009sx,Tikhonov:2009jq}.
%
%Warm dark matter can help by reconciling the observed low mass subhalos of the Milky Way~\cite{Lovell:2011rd}. A recent review of these issues is given in~\cite{Weinberg:2013aya}. There is controversy on whether warm dark matter can solve these problems, as in~\cite{Viel:2013fqw}. In~\cite{Lovell:2013ola}, a possible solution to the too big to fail problem is presented involving warm dark matter.
%
%
%}
%
%***************************

\section{Entropy in the causal diamond}

In this section we present a high level overview of the calculation of
the entropy produced in the causal diamond and how to calculate the
likelihood of a given universe using the prior probability. This
section also serves as an outline for later sections which describe our
calculations in more detail.

%To calculate the number of annihilations of DM, we start with the Boltzmann equation for annihilation.
%
%\begin{equation}
%\frac{dn}{dt} + 3 H n =  -\langle \sigma v \rangle  \left( n^2-n_{eq}^2 \right) ,
%\end{equation}
%where $\frac{dn}{dt}$ is the time rate of change of the number density
%of dark matter particles, $H$ is the Hubble parameter, $\langle \sigma
%v \rangle$ is the averaged product of annihilation cross section and
%velocity for the dark matter, and $n_{eq}$ is the equilibrium dark
%matter number density. 
%
%Most of the annihilation will come from the dense inner regions of
%dark matter halos. These will form much later than the dark matter
%freeze out so there will be effectively zero dark matter production
%and we can ignore the $n_{eq}^2$ term. We use this equation to only
%count annihilations, and not calculate the actual evolution of the
%number density, so we drop the Hubble expansion term as well: {\tt AA:
  %The previous discussion, especially the previous sentence is quite
  %confusing to me.  A couple of issues   are on my mind:  A) Dark
  %matter number density in dense cores does not obey eqn (1) (it does
  %not experience cosmic expansion). B) Regions that do experience
  %expansion {\em should} have the $3Hn$ term in because the cosmic
  %expansion affects $n$ which affects the rate.  I think the previous
  %paragraph should be rewritten to address these questions.}
%
%\begin{equation}
%\frac{dn}{dt} = -\langle \sigma v \rangle  \left(
%\frac{\rho}{m_{\chi}} \right)^2.
%\end{equation}
%Integrating over volume gives

Dark matter can annihilate when two dark matter particles find each other. As such, the dark matter annihilates with greater probability in regions of larger dark matter density. The equation that governs this interaction is given in~\cite{taylor2003} as

\begin{equation}
\frac{dN_{ann}}{dt} = -\frac{\langle \sigma v \rangle }{m_{\chi}^2} \int_V  \rho^2 dV ,
\end{equation}
where $V$ is the volume, which when applied to the CEP is taken to be the volume of the causal diamond at a given
time slice, $\rho$ is the mass density, $N_{\rm ann}$ is the total
number of dark matter particles that annihilate within that volume, and $m_{\chi}$ is the mass of the dark matter particle. 

For the entropy production from dark matter annihilations,

\begin{eqnarray}
S = g_s N_{\rm ann} ,
\end{eqnarray}
where $S$ is the total entropy production is the causal diamond, $g_s$
is the entropy increase per annihilation, and $N_{\rm ann}$ is the
total number of dark matter particles that annihilate within the
causal diamond. 

The total entropy production within the casual diamond is then

\begin{eqnarray}
S = g_s \int \int{\frac{\langle \sigma v \rangle }{m_{\chi}^2} \rho^2 d V }d t\\
S = g_s \int \int{\frac{\langle \sigma v \rangle }{m_{\chi}^2} \rho^2 \frac{d V}{d N} \frac{d N}{d M} d M} d t\\
S = g_s \int \int{\frac{\langle \sigma v \rangle }{m_{\chi}^2} \rho^2 \frac{M}{\rho_{\rm halo}} \frac{d N}{d M} d M} d t ,
\end{eqnarray}
where $N$ is the number of dark matter halos, $M$ is the mass of the halo, $\rho_{\rm halo}$ is the average density of the dark matter
halo, and the halo is taken to end at its virial radius.

%The volume per halo is $\frac{d V}{d N}$, for halos of a given mass, $M$. So, we can simplify this to give the entropy,
%
%\begin{eqnarray}
%S = g_s \int \int{\frac{\langle \sigma v \rangle }{m_{\chi}^2} \rho^2 \frac{M}{\rho_{\rm halo}} \frac{d N}{d M} d M} d t ,
%\end{eqnarray}
%where $\rho_{\rm halo}$ is the average density of the dark matter
%halo, where the halo is %assumed 
%taken to end at its virial radius.

Defining $\Delta \equiv \frac{\rho_{ \rm halo}}{\rho_m}$, where $\rho_m$ is the average matter density in the universe, we obtain

\begin{eqnarray}
S = g_s \int \int{\frac{\langle \sigma v \rangle }{m_{\chi}^2} \rho^2
  V \frac{1}{\Delta} \frac{d F}{d M} d M} d t \label{SdMdt},
\end{eqnarray}
where $F$ is the total fraction of the causal diamond's mass in halos of mass greater than $M$.
%Here for the sheth tormen fraction, if it is not in a halo, the delta
%factor is assumed to be 1, so it works out with our definition of $f$
%later on?????????
Equation~\ref{SdMdt} can be rewritten as 
\begin{eqnarray}
S = g_s \int \int{\frac{\langle \sigma v \rangle }{m_{\chi}^2} \rho_m^2 f V_{\rm com} a^{3} \frac{d F}{d M} d M} d t\\
\rho_m = \frac{3 H_0^2 \Omega_{m,0} a^{-3}}{8 \pi G}\\
f \equiv \frac{\int \rho^2 d V_{\rm halo}}{\Delta V_{\rm halo} {\rho_m}^2},
\end{eqnarray}
where $a$ is the cosmic scale factor, $V_{\rm com}$ is the comoving volume of the causal
diamond at fixed time, $f$ is the ''flux factor'', and $V_{\rm halo}$ is the volume of a halo of mass M.

%The ``flux factor''  $f$ will
%be defined in Sec.~\ref{sec:linear}.

%{\tt AA: A general comment (something I had to fix in the next
  %paragraph, as well as other places:  Think of equations (even
  %display equations) as element of a sentence and punctuate the whole
  %thing accordingly (I see you do a lot of that right).  But please
  %make sure not to start a sentence with an equation. Also note that In LaTex a line break creates a
  %new paragraph.}

According to the causal entropic principle, the probability of a parameter
having a certain value is proportional to the entropy produced in the
causal diamond multiplied by the prior probability on that parameter.
For example, taking the parameter in question to be the cosmological
constant $\Lambda$ gives 
\begin{equation}
P \propto P(U | \Lambda) P(\Lambda) ,
\end{equation}
where $P$ is the probability of observing a cosmological constant $\Lambda$. The quantity $P(\Lambda)$ is the prior
probability on $\Lambda$, and $P(U | \Lambda) = S$ is the probability
of observing a universe, given a specific value of $\Lambda$, which is
equal to the weight factor, which within the causal entropic principle
is the total entropy produced in the casual diamond.

Following~\cite{bousso2007} we take
\begin{equation}
\frac{d P(\Lambda)}{d \log \Lambda} = \Lambda \frac{d P(\Lambda)}{d \Lambda} ,
\end{equation}
where $\frac{d P(\Lambda)}{d \Lambda}$ is a constant.
So, 
\begin{equation}
\frac{d P}{d \log \Lambda} \propto \Lambda S.
\end{equation}.

%In general, the velocity averaged cross section $ \langle \sigma v
%\rangle $ can depend on velocity.
%If there was,
%{\tt AA: There is something odd about
  %this discussion:  The result of averaging cannot depend on a
  %specific value of the thing being averaged in the way this is
  %written.  Perhaps you intended to write $c/\langle v \rangle$
  %below? (and if you did, be careful about $v$ being the *magnitude*
  %of the velocity...) Either way I think you will see what I mean here,
  %and please just find some way of expressing that is more clear and consistent}
	%for instance $ \langle \sigma v \rangle \propto c / \langle v \rangle $
%For instance in the case of Sommerfeld
%enhancement, we would see greater annihilation in general, as well as more in low velocity regions. 
%But in this paper we assume that
%the velocity averaged cross section $ \langle \sigma v \rangle $ 
%is independent of velocity. This will simplify the
%calculations. Specifically,  we can take $ \langle \sigma v \rangle $
%outside the integrals giving

In this paper we assume that
the velocity averaged cross section $ \langle \sigma v \rangle $ 
is independent of the halo mass M. This need not be true in general, and specifically for the case of Sommerfeld enhancement, halos with low velocity dispersion will have boosted annihilation rates. However, our assumption will simplify the
calculations, and we take $ \langle \sigma v \rangle $
outside the integrals giving
\begin{equation}
P \propto \Lambda S = \Lambda g_s \frac{\langle \sigma v \rangle
}{m_{\chi}^2} \left(\frac{3 H_0^2 \Omega_{m,0}}{8 \pi G}\right)^2 \int
\int{f V_{\rm com} a^{-3} \frac{d F}{d M} d M} d t. 
\end{equation}
The details of how to evaluate this expression are given in
Sec.~\ref{sec:calc}. This expression is valid for the case in which
the quantity of dark matter that has annihilated is small compared to
the total quantity of dark matter. The final expression in the more
general case even when the quantity of dark matter that has annihilated grows
large is presented in Sec.~\ref{sec:total} and is the expression
used in all numeric computations. 

\section{Calculation details \label{sec:calc}}

In this section we cover the details of the assumptions and models
used to calculate the total entropy produced within the causal
diamond. Most of these are standard treatments, with the exception of
Secs.~\ref{sec:total} and \ref{sec:gs}. 

\subsection{The causal diamond \label{sec:scale}}

In the CEP, only annihilations that occur within the causal diamond
will contribute to the entropy relevant for predicting parameters in
our universe. Thus we must determine the volume of the causal
diamond. 

The first step is to find the scale factor as a function of time. We utilize the prescription for normalizing the scale factor
from Bousso et al.~\cite{bousso2007}, which requires the scale
factor to be the same for all choices of cosmological constant in the
limit of early time when the cosmological constant is negligible. We
use a set of cosmological parameters consistent with the latest
results from Planck~\cite{Ade:2013zuv}. 
 %{\tt AA: Cite something here}
Furthermore, we rescale the scale factor, $a$, so
that it is equal to one for the current value of $\Omega_{\Lambda}$,
$\Omega_{\Lambda, 0} \approx 0.68$. We assume a flat universe containing only matter and the
cosmological constant. We also use $\Omega_{m,0} = 0.32$, $\Omega_B =
0.049$, $n_s = 0.96$, and $\sigma_8 = 0.83$. 
The above choices give
\begin{equation}
a = \frac{\sqrt[3]{\Omega_{m, 0}} \sinh ^{\frac{2}{3}}\left(\frac{3}{2} \sqrt{\Omega_{\Lambda}} H_0 t\right)}{\sqrt[3]{\Omega_{\Lambda}}}.
\end{equation}
where the value of the Hubble
constant today is $H_0 = 100 h \frac{{\rm km/s}}{{\rm Mpc}}$, where $h \approx
0.67$.
In this form, what we call $\Omega_{\Lambda}$ can be greater than 1,
it is merely proportional to $\Lambda$ and in our universe matches our
value of $\Omega_{\Lambda}$. 
%{\tt AA: Is the feature mentioned in the
  %previous sentence used elsewhere in the literature, not just by us?}
This is acceptable since we cannot define a current time in any universe other than our
own. We choose this to make sense only for our universe because it
makes it more expedient to extrapolate data. An equivalent and more conceptually
pleasing form (that does not have a quantity $\Omega_\Lambda > 1$) is
\begin{equation}
a = \left(\frac{100 {\rm \frac{km/sec}{Mpc}} \sqrt{3 \omega_{m, 0}}
  \sinh \left(\frac{1}{2} \sqrt{3 \Lambda}
  t\right)}{\sqrt{\Lambda}}\right)^{\frac{2}{3}}. 
\end{equation}
where $\omega_{m, 0} = \Omega_{m, 0} h^2$ is the fraction of the
critical density comprised of matter density today multiplied by
$h^2$.

%, and $\Lambda \approx 10^{-123}$ is the value of the
%cosmological constant in Planck units. 

%Now we can continue on to the comoving volume of the causal diamond. To find this, first we find the conformal time, $\tau$.
To find the comoving volume of the causal diamond, first we find the conformal time, $\tau$.
\begin{eqnarray}
\tau = \int_0^t{\frac{d t}{a}} = \int_0^a \frac{d a}{a^2 H}\\
\tau = -\frac{_2F_1\left(\frac{1}{3},\frac{1}{2};\frac{4}{3};-\frac{\Omega_{m, 0}}{a^3 \Omega_\Lambda}\right)}{a H_0 \sqrt{\Omega_\Lambda}}
\end{eqnarray}
where $_2F_1$ denotes the Gaussian hypergeometric function.

Next we calculate the radius $r$ of the sphere enclosing the causal
diamond at a given time slice, and then hence the comoving volume of
the causal diamond, $V_{\rm com}$ on that slice:
\begin{eqnarray}
\Delta \tau = \tau(\infty) - \tau(0) = \frac{\Gamma
  \left(\frac{1}{6}\right) \Gamma \left(\frac{4}{3}\right)}{\sqrt{\pi
  } \sqrt{\Omega_\Lambda} \sqrt[3]{\frac{\Omega_{m,
        0}}{\Omega_{\Lambda }}}}\\ 
r = \frac{\Delta \tau}{2} - \left| \tau + \frac{\Delta \tau}{2} \right|\\
V_{\rm com} = \frac{4 \pi}{3} r^3(\tau)
\end{eqnarray}
\begin{figure}
\includegraphics[width=\linewidth]{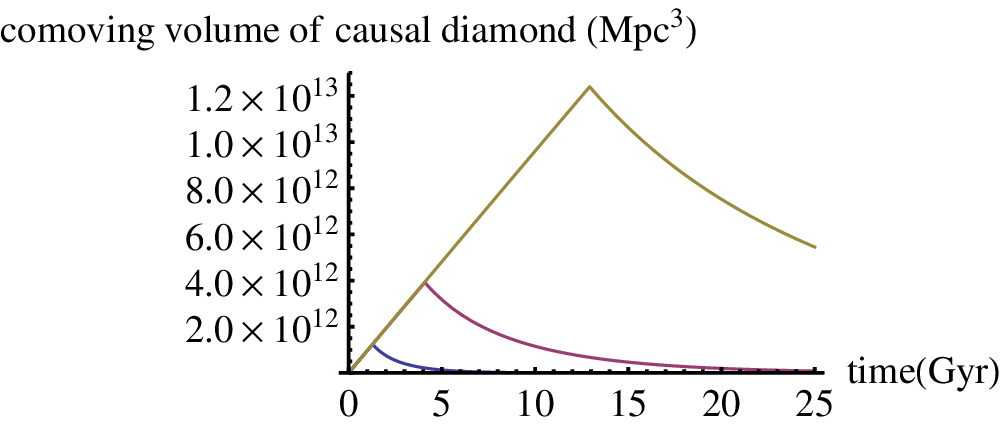}
\caption{Comoving volume of the causal diamond
  as a function of time for various choices of
  $\Lambda$. The top yellow line is for $\Lambda = 10 \Lambda_0$, the middle red line is for $\Lambda = \Lambda_0$, and the bottom blue line is for $\Lambda = 0.1 \Lambda_0$. \label{fig:vcom}} 
\end{figure}
We show the volume of the causal diamond as a function of time for
various choices of the cosmological constant in
Fig.~\ref{fig:vcom}.

\subsection{Power spectrum of the dark matter}

In order to annihilate and produce entropy, dark matter particles must
first find each other. To calculate how likely this is, we must review
how dark matter clusters together. This is measured by the matter
power spectrum. 

For generality, we use the power spectrum for warm dark matter as given in Zentner and
Bullock~\cite{zentner2003} using the BBKS transfer function~\cite{bardeen1986}. We also employ some approximations from Tegmark
et al.~\cite{tegmark2006}. The cold dark matter power spectrum
appears as a special case of this form. 
 %{\tt AA: Does the previous
  %sentence correctly capture how you are thinking about this?  The way
  %it was in the previous draft readers might have had the impression
  %that we wanted only warm dark matter, so I made some tweaks}

The root mean square (rms) matter fluctuation within a sphere of radius $R$, with a top
hat window function, $W(k R)$, is given by $\sigma(M) =
\sqrt{\sigma^2(M)}$ where 
\begin{eqnarray}
W(k R) = \frac{3}{(k R)^3}(\sin (k R)-(k R) \cos (k R))\\
\sigma^2(M) = \frac{1}{2 \pi^2}  \int_0^{\infty } \left(W(k R)\right)^2\text{P}(k) k^2 \, dk \\
R = \sqrt[3]{\frac{M}{10^{17} {\rm M_{\odot}}}} 60 h^{-1} {\rm Mpc}.
\end{eqnarray}
and $P(k)$ is the matter power spectrum, $M_\odot$ is the mass of the
sun, $M$ is the mass of the dark matter halo, and the values of
$10^{17} {\rm M_{\odot}}$ and $60 h^{-1}$ Mpc come from WMAP + SDSS
data~\cite{tegmark2006}. 
These fluctuations are given for the present time and assuming that
the growth of perturbations is linear. As large scale structure forms,
the perturbations become highly nonlinear, and will be modeled using
the halo formalism in Sec.~\ref{sec:halo}. 

The warm dark matter power spectrum is given by $P_{\rm WDM}(k) =
P_{\rm CDM}(k) \exp \left(-k{\rm R_f}-(k{\rm R_f})^2\right)$, where
$P_{\rm CDM}(k)$ is the cold dark matter power spectrum and ${\rm
  R_f}$ is the free streaming scale as given in Zentner and Bullock~\cite{zentner2003} at which the dark matter particle begins to
suppress the growth of structure:
\begin{eqnarray}
{\rm P}_{\rm CDM}(k) = 2 \pi ^2 \left(\frac{c}{H_0}\right)^{3+n} (\delta_H)^2 k^n T(q(k))^2\\
{\rm R_f}(m_{\chi}) = 0.11 {\rm Mpc} \sqrt[3]{\frac{\Omega_\chi h^2}{0.15}} \left(\frac{m_{\chi}}{{\rm keV}}\right)^{-4/3}\\
\Omega_{\chi} = \Omega_{m,0}-\Omega_B.
\end{eqnarray}
Here $ n \approx 0.96 $ is the spectral index, $\delta_H \approx 2
\times 10^{-5}$ is the fluctuation amplitude at horizon crossing,
$T(q(k))$ is the BBKS transfer function~\cite{bardeen1986}, $m_{\chi}$ is the mass of the dark matter particle, $\Omega_{\chi}$ is the
fraction of mass density in dark matter compared to the critical
density, and $\Omega_B \approx 0.049$ is the fraction of mass density
in baryons compared to the critical density: 
\begin{eqnarray}
\begin{split}
T(q)=\frac{\log (1+2.34 q)}{2.34 q} \times \\ 
\left(1+3.89 q+(16.1 q)^2+(5.46 q)^3+(6.71 q)^4\right)^{-\frac{1}{4}}
\end{split} \\
q(k)=\frac{k}{\Omega_{m,0} h^2 \exp \left(-\Omega_B-\frac{\Omega_B}{\Omega_{m,0}}\right)}
\end{eqnarray}
\begin{figure}

\includegraphics[width=\linewidth]{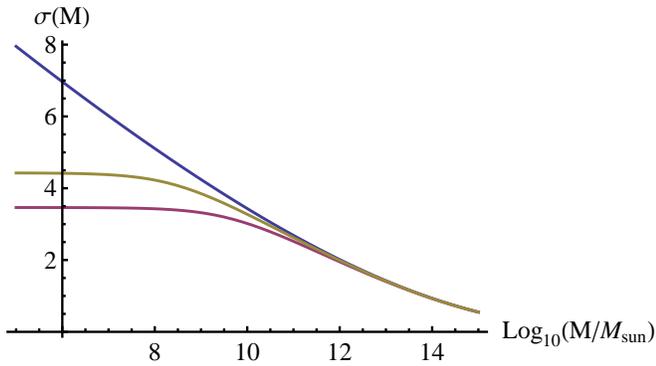}

\caption{Warm dark matter $\sigma(M)$ functions
  for two choices of the dark matter mass. The top blue line is for the
  cold dark matter model using the BBKS transfer function~\cite{bardeen1986}, and the bottom red and middle yellow lines are for dark matter
  masses of 1 and 2 keV, respectively. \label{fig:sigmas}} 

\end{figure}
Examples of these warm dark matter $\sigma(M)$ functions are shown in Fig.~\ref{fig:sigmas}.

This set of equations is valid for the rms mass fluctuations today. In
order to convert to fluctuations at other times, we simply multiply by
the growth factor of linear perturbations, $D(a)$: 
\begin{equation}
\sigma(M,a,\Lambda) = \sigma(M) D(a,\Lambda).
\end{equation}
For a flat universe consisting of only matter and a cosmological constant, the growth factor has an analytic solution 
\begin{equation}
D(a,\Lambda) = a \frac{\,
  _2F_1\left(\frac{1}{3},1;\frac{11}{6};-\frac{a^3
    \Omega_{\Lambda}}{\Omega_{m,0}}
  \right)}{_2F_1\left(\frac{1}{3},1;\frac{11}{6};-\frac{\Omega_{\Lambda,0}}{\Omega_{m,0}}
  \right)}. 
\end{equation}
We have normalized the growth factor, $D(a,\Lambda)$, to be equal to 1
at the present time in our universe. Specifically, $D(1,\Lambda_0) =
1$. 

For the purpose of numerical computation, we would like to have a
function for $\sigma(M)$ that does not have to be integrated every
time it is used. We use the fitting formula of Tegmark et al.~\cite{tegmark2006}, and cut off $\sigma(M)$ at a maximum value,
$\sigma_{\rm max}$, the maximum value $\sigma(M)$ attains as $M
\rightarrow 0$: 
\begin{equation}
\sigma_{\rm max} = \sqrt{\frac{1}{2 \pi^2}  \int_0^{\infty } \text{P}(k) k^2 \, dk}.
\end{equation}

The scale-dependence of $\sigma(M)$, $s(\mu)$, is approximately given by Tegmark et al.~\cite{tegmark2006} as
\begin{equation}
s(\mu)\approx \left[(9.1\mu^{-\frac{2}{3}})^\beta + (50.5\log_{10}(834 + \mu^{-\frac{1}{3}}) - 92)^\beta\right]^{1/\beta},
\end{equation}
where $\beta= -0.27$, and $\mu = \xi^2 M$, where $\xi^2 = \frac{1}{10^{17} \, {\rm M_{\odot}}}$.
%This formula makes the assumption that the primordial fluctuations are approximately scale-invariant and that massive neutrinos have no major effect~\cite{tegmark2006}.
This leads to a fit to the cold dark matter $\sigma(M)$ 
\begin{equation}
\sigma_0(M) \approx 0.2 \frac{\xi^{4/3}}{\rho_\Lambda^{1/3}}\delta_H s(\mu) ,
\end{equation}
where Planck units are used, and where $\rho_\Lambda$ is the energy density of the cosmological constant.

Modified with the cutoff for $\sigma_{\rm max}$, we obtain a good fit with the ansatz
\begin{equation}
\sigma(M) = \sigma_{\rm max} \left(1 - e^{-\left(\frac{\sigma_0(M)}{\sigma_{\rm max}}\right)^4}\right)^{1/4}.
\end{equation}

\subsection{Halo model and concentration mass relation \label{sec:halo}}

The greatest number of dark matter annihilations will occur in regions
in which the dark matter density is largest. These regions are dark
matter halos. We use the NFW model~\cite{NFW1997} for the density
profiles of the halos, and the model of Bullock et
al.~\cite{bullock2001} with the cosmology dependence of Dolag et
al.~\cite{dolag2003} for the concentration mass relation. 

The NFW profile is a widely used model for the density profiles of
dark matter halos, with a formally infinite mass which is cut off at
the virial radius, and divergent central density. The NFW profile is given by
\begin{eqnarray}
\rho(r) = \frac{\rho_s}{\frac{r}{r_s} \left(1 + \left(\frac{r}{r_s}\right)^2\right)} ,
\end{eqnarray} 
where $\rho(r)$ is the density, $r_s$ is the scale radius of the halo, and $\rho_s$ is twice the density at the scale radius.

We define the concentration parameter and the virial radius to match Bullock et al.~\cite{bullock2001}
\begin{eqnarray}
c \equiv \frac{r_{\rm vir}}{r_s}\\
\bar{\rho} = \Delta \rho_m\\
\Delta \approx 18 \pi^2 + 52.8 x^{0.7} + 16 x\\
x \equiv \frac{\Omega_{\Lambda}}{\Omega_m} = \frac{\Omega_{\Lambda}}{\Omega_{m,0}} a^3\\
x = \sinh^{2}\left(\frac{3}{2} \sqrt{\Omega_{\Lambda}} H_0 t\right) ,
\end{eqnarray} 
where $c$ is the concentration ratio of the halo, $r_{\rm vir}$ is the
virial radius of the halo, $\bar{\rho}$ is the average halo density
within the virial radius, and $\Delta$ is defined as the ratio of the
average halo density within the virial radius to the average density
of the universe which we have approximated using the prescription of
Tegmark et al.~\cite{tegmark2006}. 

Now it remains to calculate the clustering properties of the dark
matter halos. We will use the model of Bullock
et al.~\cite{bullock2001} for the concentration mass relation. Halos
are assumed to have been formed with a NFW profile and a concentration
of 3.5, when halos of 1/1000 of the mass of the halo would have been
collapsing.  

Bullock et al. define the scale factor of collapse ${a_{\rm c}}$ as the epoch at which
the typical collapsing mass,  $M_*({a_{\rm c}})$, equals a fixed  fraction $F$
of the halo mass at epoch $a$,
\begin{equation}
M_*({a_{\rm c}}) \equiv F M_{\rm vir}.
\label{eq1}
\end{equation}
Using the spherical collapse model, the scale factor $a$ at which
typical halos are collapsing is defined by   
$\sigma [M_*(a)]=1.686/D(a)$,
where $\sigma (M)$ is the $a=1$ linear rms density fluctuation on the
comoving scale encompassing a mass $M$, and $D(a)$ is the linear
growth rate.

Dolag et al.~\cite{dolag2003} consider concentration ratios in
alternative dark energy cosmologies. They find that the concentration
mass relation should be modified by introducing an extra factor of
$\frac{D(a)}{D_{\rm \Lambda CDM}(a)}.$ Using also their best fit
parameters for the model of Bullock et al.~\cite{bullock2001}, we obtain 
\begin{eqnarray}
c(M, \Lambda, a) = 3.5 \frac{a}{a_{\rm coll}} \frac{D(a,\Lambda)}{D(a,\Lambda_0)}\\
\sigma(0.001 M,a_{\rm coll}) = \delta_c,
\end{eqnarray}
where $\delta_c = 1.686$ is the linear theory overdensity of collapse. 
These equations are solved numerically, which gives the concentration
ratio for a given mass throughout the history of the universe in each
cosmology.

\subsection{Flux multiplier and linear theory before halos collapse \label{sec:linear}}

In order to model accurately the actual density squared of the entire
universe we must consider not only virialized halos but also obtain a
good model for what happens up until the time that they collapse. In
this case we use the simple spherical collapse model~\cite{gunngott}. To that end, we will calculate the flux factor corresponding to
the ratio of annihilation flux that would be produced from the clumpy density
profile of space divided by what would be produced if the universe
were of perfectly uniform density.
\begin{eqnarray}
f = \left\{
        \begin{array}{ll}
            \Delta f(c) & \quad \sigma > 1.686 \mbox{ or } \Delta f(c) < f(\theta) \\
            f(\theta) & \quad \mbox{otherwise}
        \end{array}
    \right.
\end{eqnarray}

For halos, plugging in the NFW profile, the result we need is
\begin{eqnarray}
f(c) = \frac{1}{\bar{\rho}^2 V} \int_0^{r_{\rm vir}} \rho(r)^2 4 \pi r^2 d r\\
\bar{\rho} = \int_0^{r_{\rm vir}} \rho(r) 4 \pi r^2 d r\\
f(c) = \frac{c^3-c^3/(1+c)^3}{9(\ln(1+c)-c/(1+c))^2} ,
\end{eqnarray} 
where $c$ is the concentration ratio of the halo, $\bar{\rho}$ is the average density of the halo within the virial radius, and $f(c)$ is the flux multiplier of~\cite{taylor2003}.

For clumps of matter that have yet to virialize and collapse, we have that~\cite{gunngott}
\begin{eqnarray}
f(\theta) = 1+\left(\frac{9 (\theta-\sin (\theta))^2}{2 (1-\cos (\theta))^3}-1\right)^2\\
\left[\left(\frac{20}{3} \sigma \right)^{3/2} = 6 (\theta-\sin (\theta)) \right] \label{eq:theta},
\end{eqnarray}
where $\theta$ is defined by the solution to \eqref{eq:theta}, and $\sigma = \sigma(M,a,\Lambda)$ is the rms matter fluctuation within a sphere of density $\rho_m$ that would enclose the mass of the clump.

\subsection{Sheth-Tormen formalism \label{sec:sheth}}

In order to predict how much dark matter annihilation occurs in halos,
we need to know how many halos there are for each mass of halo. To do
this we utilize the Sheth-Tormen model for obtaining the fraction of
halos of a given mass~\cite{sheth2001}  
%{\tt AA: Just realized: you
  %should go through and make sure each cite command is connected to the
  %previous word with a ``~''.  This prevents line breaks
  %between the word and the citations.  Please go through and check
  %that for all instances.}
\begin{equation}
\nu\,f(\nu) = 2A\,\left(1 + \frac{1}{\nu{^{2q}}}\right)\ 
\left(\frac{\nu{^2}}{2\pi}\right)^{1/2}\,\exp\left(-\frac{\nu{^2}}{2}\right) \label{eq:sheth},
\end{equation}
where $f(\nu)$ is the distribution function for the fraction of the
total mass contained in halos of mass greater than $M$, $\nu \equiv
1.686/ \sigma $, $\sigma = \sigma(M, a, \Lambda)$, $q=0.3$ and
$A\approx 0.3222$. 

The quantity that will be useful for our purposes is the fraction $d F$ of the total mass in halos with mass between $M$ and $M + d M$
\begin{eqnarray}
\frac{d F}{d M} = \frac{d F}{d \nu} \frac{d \nu}{d \sigma} \frac{d \sigma}{d M}\\
\frac{d F}{d \nu} = A\,\left(1 + \frac{1}{\nu{^{2q}}}\right) \left(\frac{2}{\pi} \right)^{1/2}\,\exp\left(-\frac{\nu{^2}}{2}\right)\\
\frac{d \nu}{d \sigma} =  - \frac{1.686}{\sigma^2} ,
\end{eqnarray}
where $\frac{d \sigma}{d M}$ is the derivative of $\sigma(M, a,
\Lambda)$ with respect to $M$.

%{\tt AA: Clarify what ``here'' means.  Do I infer correctly that
  %it means ``not low mass cutoff for $\sigma$''?  If so we should be
  %clearer that that is what we are taking... or is it that CDM implies
  %something about $\sigma_{max}$?}

This formalism works well if we assume that there is no free streaming scale for the
dark matter, so that it has the property that all the mass is contained in
halos of some size:
\begin{eqnarray}
\int_0^\infty \frac{d F}{d M} d M = 1.
\end{eqnarray}
But with a cutoff for sigma at low mass scales this no longer works because the rms mass fluctuations cannot go to infinity at small mass scales:
\begin{eqnarray}
\int_0^\infty \frac{d F}{d \sigma} d \sigma = 1\\
\int_0^{\sigma_{\rm max}} \frac{d F}{d \sigma} d \sigma \neq 1.
\end{eqnarray}
However, the Sheth-Tormen formalism has been tested and fit by data
from large structures in our own universe, where there is no
difference between the warm dark matter clustering and the cold dark
matter clustering because the matter power spectrum is not yet cut off
at those large mass scales. So there is a quantity of matter that
simply does not collapse into halos, which we can just use the linear
theory results to account for. This becomes more and more the case at
earlier times. There is also an effect to be accounted for at large
mass scales, since these will never collapse into a halo, even in the
infinite future. These objects should be treated according to the
linear theory too, since they will not collapse. 
For a more detailed treatment of warm dark matter clustering, see~\cite{dunstan2011,angulo2013,schneider2013}.

%Schneider et. al. actually ignore this issue in their paper, but mention it in the conclusions as an unverified assumption that could cause their extended Press Schechter method to break down completely, but Smith and Markovic \cite{Smith:2011ev} deal with this issue, as does Benson et. al. \cite{Benson:2012su}, however their predictions differ considerably from each other by orders of magnitude \cite{angulo2013}.

For our purposes, we are concerned with the largest sources of
annihilation, which are those dense halos which annihilate rapidly,
for which there is no conflict. This will in fact be overwhelmingly
larger than the contribution from annihilations in regions that do not
clump to form halos. At early times when the smooth not yet clumped
together portion becomes large, the total amount of annihilation in
the causal diamond becomes very small. Thus we expect our neglect of
this issue to have little effect on the results.

\subsection{Total annihilation \label{sec:total}}

Up until this point, we have assumed that the annihilation of dark
matter does not have any feedback effect on the density profiles of
halos. But as more and more dark matter annihilates at late times,
eventually so much will have annihilated that it will become an
appreciable fraction of the total dark matter particles available to
annihilate and annihilation will slow down. This is very important
since it cuts off the otherwise potentially (formally) unbounded amount of
annihilation at late times. 

We model this in a very simple way by assuming that for each mass
scale of halo there is some maximum number of annihilations that can
occur, and that furthermore, these annihilations modify the halo
density profile only by changing the total overall density, and do not
heat the halo or slow down the annihilation faster in the more dense
central regions where more annihilations occur as would be the case
for real halos. 

We begin with the Boltzmann equation for annihilation,

\begin{eqnarray}
\frac{d n}{d t} + 3 H n = - \langle \sigma v \rangle \left( n^2 - n_{eq}^2 \right) ,
\end{eqnarray}
where $n$ is the number density of dark matter particles, and $n_{eq} \approx 0$ is the equilibrium number density of dark matter.

On scales of mass $M$,
\begin{eqnarray}
\frac{d N}{d t} = -\frac{\langle \sigma v \rangle }{m_{\chi}^2} \left(\frac{3 H_0^2 \Omega_{m,0}}{8 \pi G}\right)^2 f a^{-3} \left(\frac{N}{N_{\rm max}}\right)^2\\
N_{\rm max} = \frac{\rho_m}{m_\chi} a^3 = \frac{\rho_{m,0}}{m_\chi} ,
\end{eqnarray}
where $N = n a^3$, and $N_{\rm max}$ is the initial value of $N$, before annihilations from clustering decrease the number density of dark matter particles.

%And so, with this simplified model 
%\begin{eqnarray}
%\frac{d N}{d t} = \frac{\langle \sigma v \rangle }{m_{\chi}^2} \left(\frac{3 H_0^2 \Omega_{m,0}}{8 \pi G}\right)^2 \int{f V_{\rm com} a^{-3} \frac{d F}{d M} d M} .
%\end{eqnarray}

%On scales of mass $M$, we can plot $\frac{d N}{d t}$ to determine its time dependence:
%\begin{eqnarray}
%\frac{d n(M) a^3}{d t} = -\frac{\langle \sigma v \rangle }{m_{\chi}^2} \left(\frac{3 H_0^2 \Omega_{m,0}}{8 \pi G}\right)^2 f a^{-3}
%\end{eqnarray}
%Empirically, we determine that 
%\begin{eqnarray}
%\frac{d n(M) a^3}{d t} \propto -e^{\alpha t}
%\end{eqnarray}

\begin{figure}
\vspace{20pt}
\includegraphics[width=\linewidth]{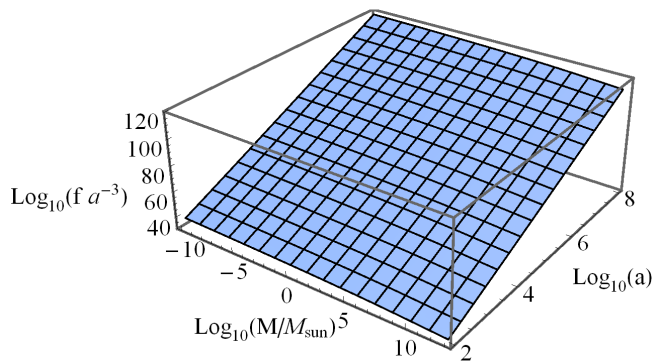}
\caption{$\log(f a^{-3})$ vs. $\log(a)$ for the warm
  dark matter for $\Omega_\Lambda = 0.7$ and $h = 0.7$ and $\Omega_m =
  0.3$. This surface can be well approximated by the linear fit
  $\log(f a^{-3}) = 13.4 \log(a) + 20.7$. The slope of this fit is the
  same across all choices of parameters and is valid for all mass
  scales that eventually collapse into halos, in our model, for $\log
  \left(\frac{M}{M_\odot}\right) < 13.6$.\label{fig:total}} 
	
\end{figure}

We plot $\log(f a^{-3})$ vs. $\log(a)$ in Fig.~\ref{fig:total} and find
\begin{eqnarray}
\log(f a^{-3}) = \beta \log(a) + C,
%f a^{-3} = e^C a^{\beta}\\
%\frac{d N(M)}{d t} \propto a^{\beta} ,
\end{eqnarray}
where $\beta \approx 13.4$, and we are far into cosmological constant
domination by the time all the dark matter has annihilated. So, 
\begin{eqnarray}
a \propto e^{\sqrt{\Omega_\Lambda} H_0 t}\\
\frac{d N}{d t} = -\langle \sigma v \rangle f a^{-3} N^2\\
%\frac{d N}{d t} = -\langle \sigma v \rangle e^{C+\sqrt{\Omega_\Lambda} H_0 \beta t} N^2\\
\frac{d N}{N^2} = -\langle \sigma v \rangle e^{C+\sqrt{\Omega_\Lambda} H_0 \beta t} d t\\
%\frac{1}{N} = C_1 + \frac{\langle \sigma v \rangle}{\sqrt{\Omega_\Lambda} H_0 \beta} e^{C+\sqrt{\Omega_\Lambda} H_0 \beta t}\\
%\frac{1}{N} = C_1 + \frac{\langle \sigma v \rangle}{\sqrt{\Omega_\Lambda} H_0 \beta} f a^{-3}\\
\frac{1}{N} = \frac{1}{N_{\rm max}} + \frac{\langle \sigma v \rangle}{\sqrt{\Omega_\Lambda} H_0 \beta} f a^{-3}
%\frac{N}{N_{\rm max}} = \frac{1}{1 + \frac{\langle \sigma v \rangle}{\sqrt{\Omega_\Lambda} H_0 \beta} f a^{-3} N_{\rm max}}
\end{eqnarray}

We then adjust the expression for entropy production within the causal diamond by the ratio of $\left(\frac{N}{N_{\rm max}}\right)^2$, yielding the final form to go into the numerical computation, 
\begin{widetext}
\begin{eqnarray}
%\frac{d N}{d t} = \frac{\langle \sigma v \rangle }{m_{\chi}^2} \left(\frac{3 H_0^2 \Omega_{m,0}}{8 \pi G}\right)^2 \int{f V_{\rm com} a^{-3} \frac{d F}{d M} \frac{N_{\rm max}^2 }{\left(N_{\rm max} + \frac{1}{\alpha} \left(\frac{d N(M)}{d t}\right)_0 \right)^2} d M}\\
%\frac{d N}{d t} = \frac{\langle \sigma v \rangle }{m_{\chi}^2} \left(\frac{3 H_0^2 \Omega_{m,0}}{8 \pi G}\right)^2 \int{f V_{\rm com} a^{-3} \frac{d F}{d M} \frac{N_{\rm max}^2 }{\left(N_{\rm max} + \frac{1}{\alpha} \frac{\langle \sigma v \rangle }{m_{\chi}^2} \left(\frac{3 H_0^2 \Omega_{m,0}}{8 \pi G}\right)^2 f a^{-3} \right)^2} d M}\\
%P \propto S \Lambda = \Lambda g_s \frac{\langle \sigma v \rangle }{m_{\chi}^2} \left(\frac{3 H_0^2 \Omega_{m,0}}{8 \pi G}\right)^2 \int \int{f V_{\rm com} a^{-3} \frac{d F}{d M} \frac{N_{\rm max}^2 }{\left(N_{\rm max} + \frac{1}{\alpha} \frac{\langle \sigma v \rangle }{m_{\chi}^2} \left(\frac{3 H_0^2 \Omega_{m,0}}{8 \pi G}\right)^2 f a^{-3} \right)^2} d M} d t\\
%P \propto S \Lambda = \Lambda g_s \frac{\langle \sigma v \rangle }{m_{\chi}^2} \left(\frac{3 H_0^2 \Omega_{m,0}}{8 \pi G}\right)^2 \int \int{f V_{\rm com} a^{-3} \frac{d F}{d M} \frac{1}{\left(1 + \frac{1}{\alpha} \frac{\langle \sigma v \rangle }{m_{\chi}} \left(\frac{3 H_0^2 \Omega_{m,0}}{8 \pi G}\right) f a^{-3} \right)^2} d M} d t\\
P \propto S \Lambda = \Lambda g_s \frac{\langle \sigma v \rangle
}{m_{\chi}^2} \left(\frac{3 H_0^2 \Omega_{m,0}}{8 \pi G}\right)^2 \int
\int{f V_{\rm com} a^{-3} \frac{d F}{d M} \frac{1}{\left(1 +
    \frac{1}{\sqrt{\Omega_\Lambda} H_0 \beta} \frac{\langle \sigma v
      \rangle }{m_{\chi}} \left(\frac{3 H_0^2 \Omega_{m,0}}{8 \pi
      G}\right) f a^{-3} \right)^2} d M} d t . 
\label{TheBigOne}
\end{eqnarray}
\end{widetext}

%\begin{widetext}
%\begin{eqnarray}
%P \propto S \Lambda = \Lambda g_s \frac{\langle \sigma v \rangle }{m_{\chi}^2} \left(\frac{3 H_0^2 \Omega_{m,0}}{8 \pi G}\right)^2 \int \int{f V_{\rm com} a^{-3} \frac{d F}{d M} \frac{1}{\left(1 + \frac{1}{\sqrt{\Omega_\Lambda} H_0 \beta} \frac{\langle \sigma v \rangle }{m_{\chi}} \left(\frac{3 H_0^2 \Omega_{m,0}}{8 \pi G}\right) f a^{-3} \right)^2} d M} 10^w \log 10 d w .
%\end{eqnarray}
%\end{widetext}

\subsection{Entropy per annihilation \label{sec:gs}}

One of the most important parameters governing how much entropy is
produced by dark matter annihilations is the amount of entropy
produced by each dark matter annihilation, $g_s$. This in general
depends on both the particle physics model for the annihilation and the astrophysical details of how the end state particles thermalize. If the
dark matter produces photons that eventually scatter off dust
particles like in the case of ordinary starlight, the entropy per
annihilation is proportional to the energy of the dark matter
particle. If the dark matter annihilates to neutrinos that do not later interact, the entropy per annihilation will be a constant.

%{\tt The dark matter could annihlate to photons, or neutrinos or various other particles. Higher energy photons might not thermalize completely, and neutrinos would not interact. So a gamma of 1 would be the likely maximum, to a gamma of zero, with a fixed entropy gain per annihilation e.g. $\chi \rightarrow \nu \nu$.}

At this stage in our understanding of dark matter, we find it more
convenient to just assume the simple functional form of a power law
for the dependence of the expected entropy produced per dark matter
annihilation as a function of the mass of the dark matter
particle\footnote{Formally, one should think of this formula as the product of the entropy
  production per particle with the prior probability assigned to a
  theory with dark matter mass $m_{\chi}$.}: 
\begin{eqnarray}
g_s = g_0 \left(\frac{m_{\chi}}{\rm keV}\right)^{\gamma} ,
\end{eqnarray}
where $g_0$ is the entropy per annihilation of a 1 keV dark matter
particle, and we have introduced $\gamma$ as the parameter determining
the dependence on mass. 
This parameterization allows us explore the predictions of which
values of dark matter mass are favored as a function of $\gamma$ and
$g_0$. 

%{\tt Higher cross section could also mean lower relic density of dark matter. We could hold the relic density fixed, and then the cross section would have dependence on the mass of dark matter. We could incorporate this with the gamma factor on the prior on dark matter mass together into a single mass dependence relationship.}

\section{Cross section}

The cross section for annihilation will affect both the initial abundance of dark matter, and the later annihilation rate. In all sections except Sec.~\ref{sec:varycross}, we hold the matter density fixed at $\Omega_m = 0.32$, in order that the fitting functions derived from observations and simulations will be accurate. In Sec.~\ref{sec:varycross}, we explore the effect of varying the cross section for dark matter annihilation on the total entropy production. We assume that the same cross section that governs annihilation in the early universe holds in the late universe as well. We solve the Boltzmann equation numerically as in~\cite{Gondolo:1990dk} to obtain the initial amount of dark matter, which will determine the baryon fraction, and the matter abundance. Then we use the relations of~\cite{cline2008} to adjust the clustering properties of matter. Note that in Sec.~\ref{sec:varycross} the fitting functions are still based on simulations of universes with values of the matter density, $\Omega_m \approx 0.32$, a weakness we hope to correct in future work.

%In this case we are assuming that the dark matter is a thermal relic and use the appropriate free streaming formulae.

\section{Numerical methods}

In this section, we present the details of how the formulas presented
earlier were processed numerically to generate the final probability
plots. 

In performing the calculations, we can calculate the differential
entropy production per unit time per unit mass of halo. It is
convenient to parameterize time and mass scale logarithmically using
\begin{eqnarray}
M = 10^y\\
t = 10^w ,
\end{eqnarray}
where for convenience, we express time in units of Gyr, and mass in units of solar masses, $M_\odot$.

Then the task is to integrate Eq.~\eqref{TheBigOne} over all time and all mass
scales. Our integrations converge nicely thanks to the intrinsically
finite physical processes which are being computed. 
Performing a 2D array of these 2D integrals to get the
output plots as a function of $\Lambda$ and dark matter mass was not
overly resource intensive and was done on a laptop.  
%{\tt AA: I
  %rewrote the previous paragraph.  Is it correct, and does it convey
  %what you wanted? Same question for the next paragraph:}

The time integral converges at the point when when most of the dark matter has already
annihilated and the rate slows dramatically. At early times the annihilation rate is very rapidly increasing with time as halos
cluster more and more. Early epochs have much
less annihilation and so ultimately it is only the final (logarithmic)
epoch just before all
the dark matter annihilates that dominates the integral.  

\begin{figure}
\includegraphics[width=\linewidth]{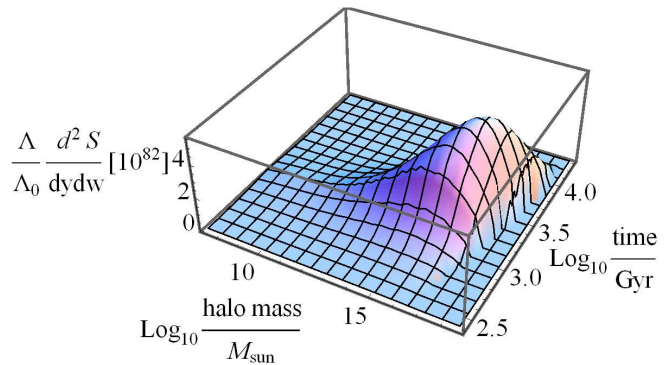}
\caption{An example of the integrand of the integral for total entropy plotted over the region in which the integral is significant. In this plot, $\Lambda = 10^{-5} \Lambda_0$ and $m_{\chi} = 1$ keV. This type of plot shows that this integral to get the total entropy is convergent and also the time and mass scales at which the greatest number of annihilations will occur. \label{fig:region}}
\end{figure}

The cutoff at low mass scales for halos occurs where the free
streaming scale of the dark matter particle cuts off structure
formation. 
%{\tt AA: What about CDM?} 
This can go down to keV ranges with our model until it cuts
off almost all structure formation on all scales. The cutoff at high
mass scales comes from the Sheth-Tormen formalism. There is a maximum
size limit to the size of bound objects that increases with time, so
at late times this cutoff is at larger mass scales than at early
times. A typical plot of the integrand in the region in which the integral is significant is shown in
Fig.~\ref{fig:region}. 
%{\tt AA: Please clarify here and in the
  %caption:  I think we are showing the actual integrand in the figure
  %(label the z axis), and the figure shows nicely that the integral is
  %convergent.  Please distinguish more clearly between the region of integration
  %and the region where the integrand is significant.}

\section{Results}

Here we present the results of our calculations. In
Sec.~\ref{sec:dmonly} we present the probability distribution
in $\Lambda$--$m_{\chi}$ space that results from only including the entropy produced by dark matter
annihilation in the CEP weighting.  This limited analysis helps us understand the impact
of adding the new (dark matter annihilation) ingredient.  In 
Sec.~\ref{sec:currentlambda} we scrutinize this distribution further
by examining a slice at fixed $\Lambda$ (corresponding to the
concordance value). We present results from adding the (usual) stellar burning contribution in
Sec.~\ref{sec:withstars}. In Sec.~\ref{sec:varycross} we explore the effect of varying the cross section for dark matter annihilation, both on the stellar entropy and the entropy from dark matter annihilation.

\subsection{Entropy in the causal diamond \label{sec:dmonly}}

\begin{figure}

\includegraphics[width=\linewidth]{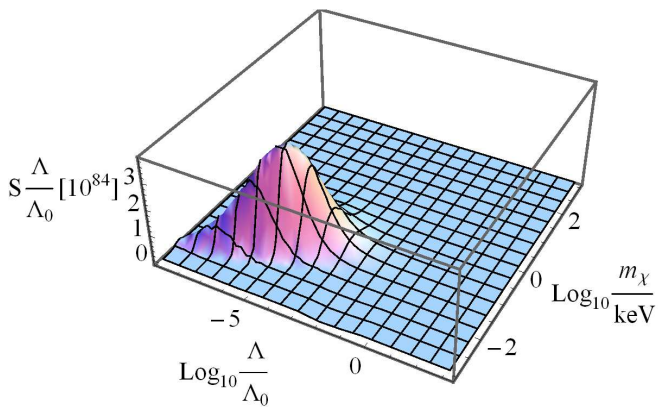}

\caption{Probability of observing a given value of the cosmological constant and mass of dark matter, assuming $g_s = 1$, $\langle \sigma v \rangle = 3 \times 10^{-26} \rm{\frac{cm^3}{s}}$, and that dark matter annihilation is the only contributor to entropy
  production in the causal diamond. The peak of this plot occurs for
  $\Lambda \approx 9.3 \times 10^{-6} \Lambda_0$ and $m_{\chi} \approx
  0.04$ keV. \label{fig:probplotDM}} 

\end{figure}

\begin{figure}

\includegraphics[width=\linewidth]{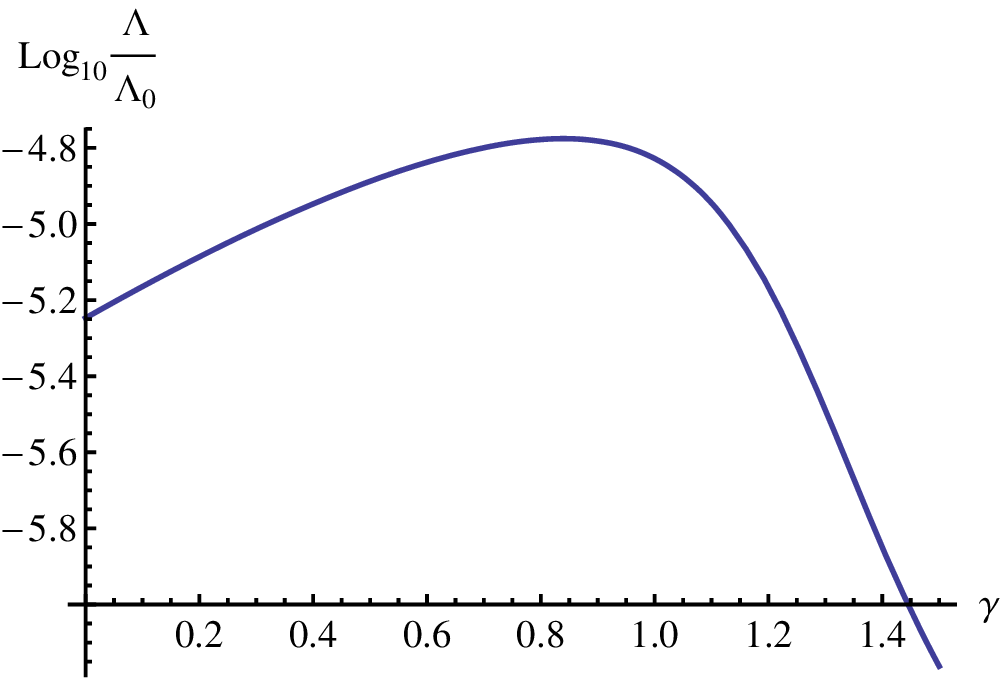}

\caption{The expected value of the cosmological
  constant marginalized over mass of dark matter as a
  function of the power law index $\gamma$ for the entropy per
  annihilation, assuming $g_s \propto {m_{\chi}}^\gamma$, and $\langle \sigma v \rangle = 3 \times 10^{-26} \rm{\frac{cm^3}{s}}$. Here only entropy production from annihilations is
  included, and we neglect entropy production from stars. \label{fig:probplotDMlambdagamma}} 

\end{figure}

\begin{figure}

\includegraphics[width=\linewidth]{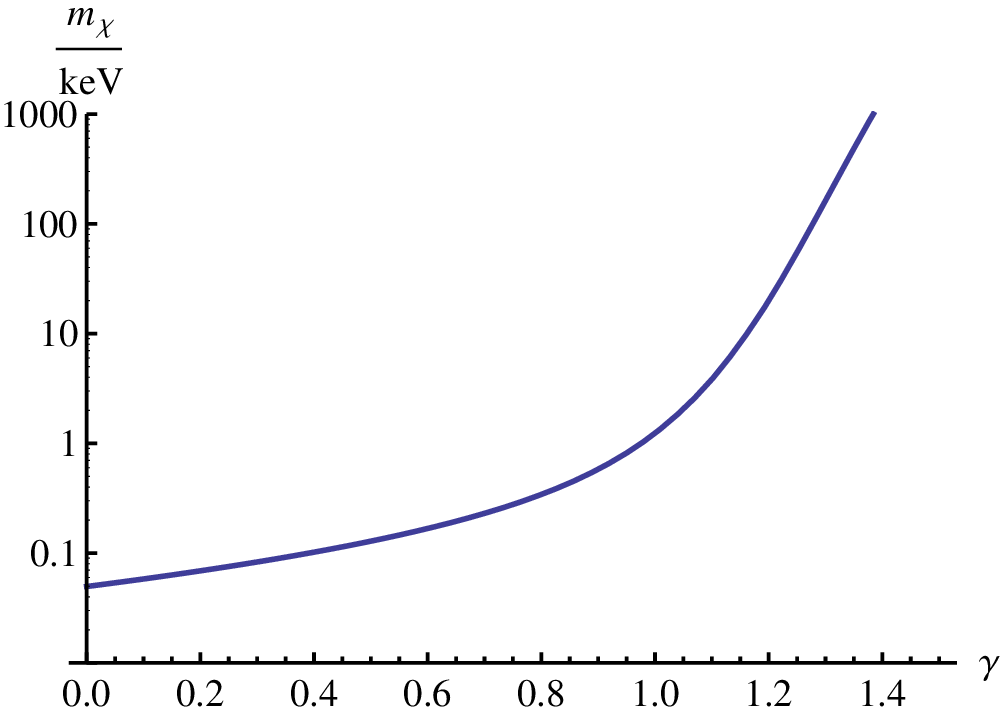}

\caption{The expected value of dark matter mass marginalized over the cosmological
  constant as a function of the power law index $\gamma$ for the entropy per
  annihilation, assuming $g_s \propto {m_{\chi}}^\gamma$, and $\langle \sigma v \rangle = 3 \times 10^{-26} \rm{\frac{cm^3}{s}}$. Here only entropy production from annihilations is
  included, and we neglect entropy production from stars. \label{fig:probplotDMmassgamma}} 

\end{figure}

\begin{figure}

\includegraphics[width=\linewidth]{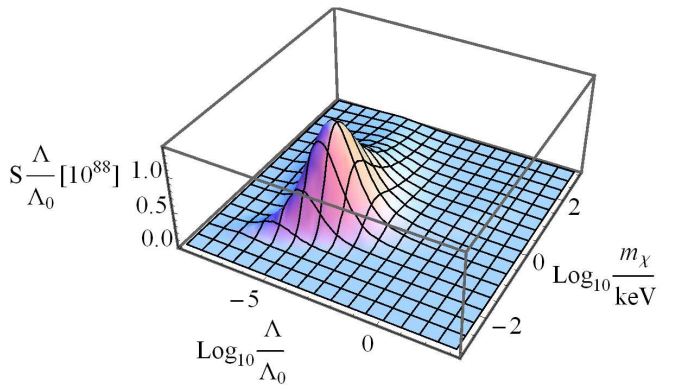}

\caption{Probability of observing a given value of the cosmological constant and mass of dark matter, assuming $g_s = \frac{m_{\chi}}{20 \rm{meV}}$, $\langle \sigma v \rangle = 3 \times 10^{-26} \rm{\frac{cm^3}{s}}$, and that dark matter annihilation is the only
  contributor to entropy production in the causal diamond. This plot assumes the dark matter annihilates into photons which are scattered by
  dust at energies of 20 meV. The peak of this plot occurs for
  $\Lambda \approx 3.2 \times 10^{-5} \Lambda_0$ and $m_{\chi} \approx
  0.14$ keV. \label{fig:probplotDMdust}} 

\end{figure}

We obtained the total entropy produced in the causal diamond, due only
to dark matter annihilations, as a function of $\Lambda$ and mass of
the dark matter. This is then weighted by the prior on $\Lambda$, which
is assumed to be proportional 
to the value of $\Lambda $ itself (following the standard approach). A plot of the probability distribution is shown in
Fig.~\ref{fig:probplotDM}. Since small values of $\Lambda$ allow the
causal diamond to stay large for a longer period of time, they will produce more
entropy than large values do. So the cutoff for $\Lambda$ occurs where the
extra size of the causal diamond at late times does not help gain much
more entropy. This typically occurs at the epoch at which a sizable
fraction of the dark matter has annihilated. This in general happens
in the far future, and leads to the low favored values for $\Lambda$,
in contrast to traditional CEP calculations (which model stellar entropy production). 

In the dark matter mass direction, low masses allow more annihilations
because there are more dark matter particles for the same total dark
matter mass,
and hence they are more likely to find each other and annihilate. When
we vary the power law index $\gamma$ for the entropy per annihilation
in Figs.~\ref{fig:probplotDMlambdagamma} and
\ref{fig:probplotDMmassgamma}, we see that a low mass dark matter
particle is favored for all $\gamma < 1.4$, and that for higher values
of $\gamma$, the dark matter is favored to have arbitrarily high mass
approaching infinite mass in our crude model with no cutoff for
mass. This is because low values of dark matter mass will produce more
entropy because their annihilation rate goes as the number density
squared, but too low values of the dark matter mass will suppress
structure formation too well and will not allow enough
annihilation-enhancing halos to form. In general the balance between these
effects favors a dark matter mass pushing up against the limit where structure formation is totally 
cut off, which happens just below the keV scale in our model. If we use the simple model that dark
matter annihilations produce light which gets most of its entropy from
heating of dust just like ordinary starlight, as in~\cite{bousso2007}, we get the plot presented in
Fig.~\ref{fig:probplotDMdust}. This model produces more peak entropy
from dark matter annihilations than would be produced by stars. 

The preferred value of the cosmological constant is roughly $10^{-5}$ times 
the measured value today, regardless of the value of $\gamma$. In general, most of the annihilation of dark
matter will occur in the far future as halos continue to collapse more
and more. 

%{\tt AA: the following sentence is confusing:  Is this
  %phenomenon considered normal?  Is it included in our model?} If the
%collapse of halos reaches a maximum and slows down in the 
%future, much less entropy will be produced in the causal diamond
%because the casual diamond will be much smaller at later times. 

\subsection{Assuming our current value for $\Lambda$ \label{sec:currentlambda}}

\begin{figure}

\includegraphics[width=\linewidth]{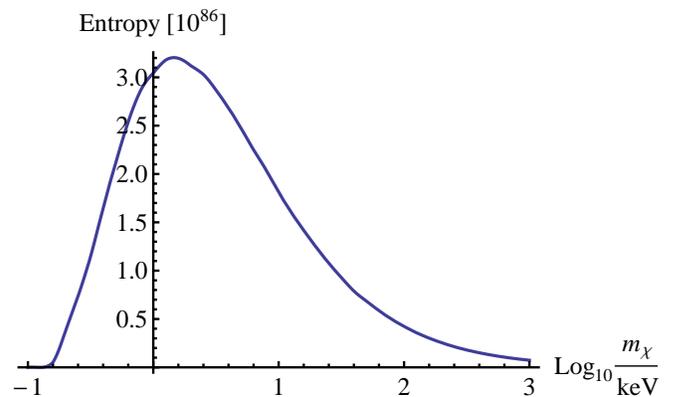}

\caption{Probability of observing a given mass of the dark
  matter particle, fixing the cosmological constant at its current measured value,
  $\Omega_{\Lambda} \approx 0.68$, assuming $g_s = \frac{m_{\chi}}{20 \rm{meV}}$, $\langle \sigma v \rangle = 3 \times 10^{-26} \rm{\frac{cm^3}{s}}$, and that dark matter annihilation is the only
  contributor to entropy production in the causal
  diamond. \label{fig:probcurrentlambda}} 

\end{figure}

\begin{figure}

\includegraphics[width=\linewidth]{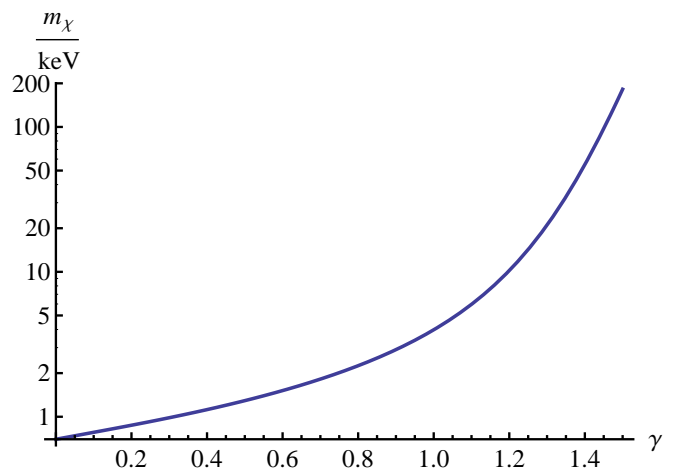}

\caption{The expected value of dark matter mass as a
  function of the power law index $\gamma$ for the entropy per
  annihilation, assuming $\Omega_\Lambda = 0.68$, $g_s \propto {m_{\chi}}^\gamma$, and $\langle \sigma v \rangle = 3 \times 10^{-26} \rm{\frac{cm^3}{s}}$. Here only entropy production from annihilations is included, and we neglect entropy production from stars. \label{fig:probcurrentlambdagamma}} 

\end{figure}

These preferred values of $\Lambda$ do not agree with the current
accepted value for $\Lambda$. 
%{\tt We should compare with other CEP
  %calculations which tolerate some discrepancy.. how much worse in our
  %case here?}
Using the standard causal entropic principle with entropy from stars predicts agreement within 1 sigma for the cosmological constant. We find a disagreement of at minimum 2 sigma in the opposite direction for all choices of dark matter mass if we assume all the entropy production comes from dark matter annihilation. We find it interesting to assume some
other consideration fixes the observed value of $\Lambda$ (perhaps
simply a prior based on current data). We then might still be able to make
useful predictions about the favored value of the dark matter mass, by
simply assuming the observed value for $\Lambda$, rather than letting
it vary. We plot the relative probability for dark matter mass
fixing $\Lambda$ to the currently measured value for the model in which energy from dark matter annihilations produces entropy from heating of dust ($\gamma=1$) in Fig.~\ref{fig:probcurrentlambda}. This peaks at around 1 keV or so,
which is an order of magnitude more massive than the peak if $\Lambda$
is allowed to vary. We can also determine the most likely value for
the dark matter mass as a function of $\gamma$ here too, which we show
in Fig.~\ref{fig:probcurrentlambdagamma}. We can see that here the
dependence on $\gamma$ is more mild than in the case where we also
allow $\Lambda$ to vary. 

%The main uncertainty here is the entropy per annihilation for different masses of dark matter, and also the dependence of the cutoff for the $\sigma(M)$ spectrum on dark matter mass. Both of these are strongly dependent on the model for dark matter, for which we have just chosen as simple a model as possible, and parametrized away most of our ignorance.

\subsection{Combining with stars \label{sec:withstars}}

\begin{figure}

\includegraphics[width=\linewidth]{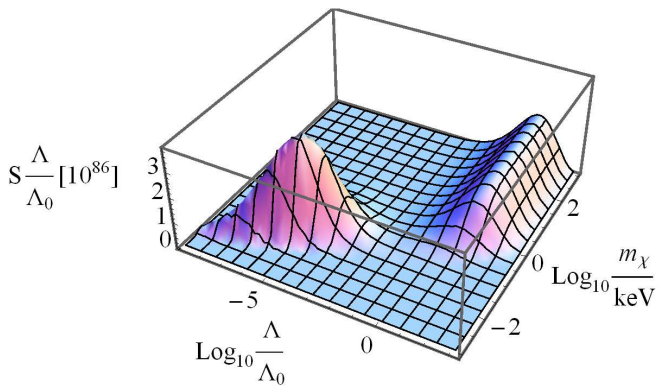}

\caption{Probability of observing a given value of the cosmological constant and mass of dark matter,
  assuming $g_s = 100$, $\langle \sigma v \rangle = 3 \times 10^{-26} \rm{\frac{cm^3}{s}}$, and that both dark matter annihilation and stars contribute to
  entropy production in the causal diamond. The relative heights of
  the two peaks depend on the parameters given to the dark matter
  model. 
	%Here, they are tuned to provide the same peak contribution. 
	\label{fig:probwithstars}} 

\end{figure}

Dark matter annihilations have the potential to produce a very large
amount of entropy within the casual diamond, but do they produce
enough entropy to compete with the entropy from dust heated by
starlight? Following~\cite{cline2008}, we add in the contribution from stars, calculated as in
Bousso et al.~\cite{bousso2007}, and using the star formation rate of
Hernquist and Springel~\cite{Hernquist:2002rg}, we can see how much
entropy per annihilation is needed to shift the predictions, and explore
the parameter space of the true solution using both entropy from stars
and dark matter annihilation. This contribution is seen to result in
two distinct peaks in the probability plot, as shown in
Fig.~\ref{fig:probwithstars}. The locations of both peaks have
little dependence on each other, since there was likely little dark
matter annihilation during the epoch of peak star formation and there
will be little star formation in the epoch where the peak of dark
matter annihilation is calculated to occur. So, the value of the
cosmological constant predicted from this combined approach is either
the value obtained as in Bousso et al.~\cite{bousso2007}, or the value
optimized by the dark matter annihilation obtained here, depending on
which regime dominates the probability.  We note that while the preferred values for $\Lambda$ found
in~\cite{bousso2007} were on the high side, and the dark matter
annihilations prefer low $\Lambda$'s, mixing the two produces a
bimodal probability distribution, not a single shifted peak.  So our
result does not offer a simple way to tune the preferred value (which
in any case is not too far off the observed value in the limit when either effect dominates). 

\subsection{Varying the cross section \label{sec:varycross}}

\begin{figure}

\includegraphics[width=\linewidth]{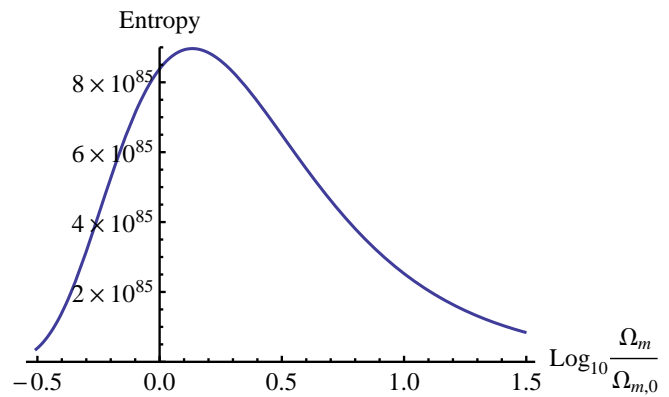}

\caption{Probability of observing a given value of the overall matter abundance, by varying the cross section of dark matter annihilation, and holding fixed baryonic physics, assuming stellar entropy dominates over entropy from dark matter annihilations in the causal diamond.\label{fig:starsigv}}

\end{figure}

We can add in the effect of varying the cross section for dark matter annihilation to the results. Firstly, if we consider the effect of varying the dark matter annihilation cross section on the entropy produced from stars, this will directly influence the total quantity of dark matter.
If we vary the quantity of dark matter directly, we see that the combined effect on both the matter abundance and baryon fraction results in a preferred value of $\Omega_m$ in excellent agreement with the observed value as shown in Fig.~\ref{fig:starsigv}. While~\cite{cline2008} vary both the matter abundance and baryon fraction separately, we get better agreement than they do when we vary the dark matter cross section holding baryonic physics fixed, which we believe is more physically motivated.

%\subsection{Filtering scale of warm dark matter}
%
%We begin with the work of~\cite{Hogan:1999hj}, where we write the filtering scale as
%
%\begin{eqnarray}
%k = \frac{H}{{\langle v^2 \rangle}^{1/2}}\\
%Q = \frac{\rho}{{\langle v^2 \rangle}^{3/2}} = 2 \times 10^{-3} g_{*} {m_\chi}^4\\
%\rho_\chi = f(\lambda) \frac{g_{*}}{\pi^2} \left(\frac{k T}{\hbar c} \right)^3 m_\chi
%\end{eqnarray}
%
%We end up with
%
%\begin{eqnarray}
%k = H \left(\frac{2 \times 10^{-3} \pi^2}{f(\lambda)}\right)^{1/3} \left(\frac{k T}{\hbar c} \right)^{-1} m_\chi\\
%k =  \frac{H c m_\chi}{k T} \left(\frac{2 \times 10^{-3} \pi^2}{f(\lambda)}\right)^{1/3}\\
%k =  \frac{m_\chi c^2}{k T} \left(\frac{2 \times 10^{-3} \pi^2}{f(\lambda)}\right)^{1/3}\left(\frac{c}{H}\right)^{-1},
%\end{eqnarray}
%
%where this in in the proper units!
%
%Using the work of~\cite{Viel:2005qj},
%
%\begin{eqnarray}
%T(k) = \left(\frac{P(k)_{\rm \Lambda WDM}}{P(k)_{\rm \Lambda CDM}}\right)^{1/2}\\
%T(k) = \left(1+(\alpha k)^{2 \nu}\right)^{-5/\nu}\\
%\alpha = 0.049\left(\frac{m_\chi}{1 \rm keV}\right)^{-1.11}\left(\frac{\Omega_\chi}{0.25}\right)^{0.11}\left(\frac{h}{0.7}\right)^{1.22} h^{-1} {\rm Mpc},
%\end{eqnarray}
%
%where $\nu = 1.12$ for $k<5h {\rm Mpc}^{-1}$, and $\alpha$ is valid for a thermal relic dark matter.

\begin{figure}

\includegraphics[width=\linewidth]{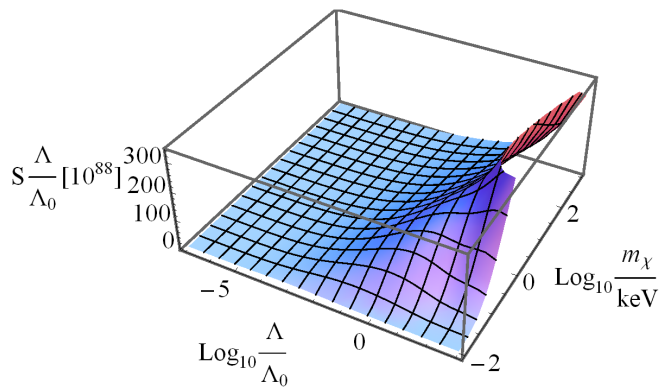}

\caption{Probability of observing a given value of the cosmological constant and mass of dark matter, assuming $g_s = \frac{m_{\chi}}{20 \rm{meV}}$, $\langle \sigma v \rangle = 3 \times 10^{-28} \rm{\frac{cm^3}{s}}$, and that dark matter annihilation is the only
  contributor to entropy production in the causal diamond. This plot assumes the dark matter annihilates into photons which are scattered by
  dust at energies of 20 meV. \label{fig:sigvneg2}}

\end{figure}

\begin{figure}

\includegraphics[width=\linewidth]{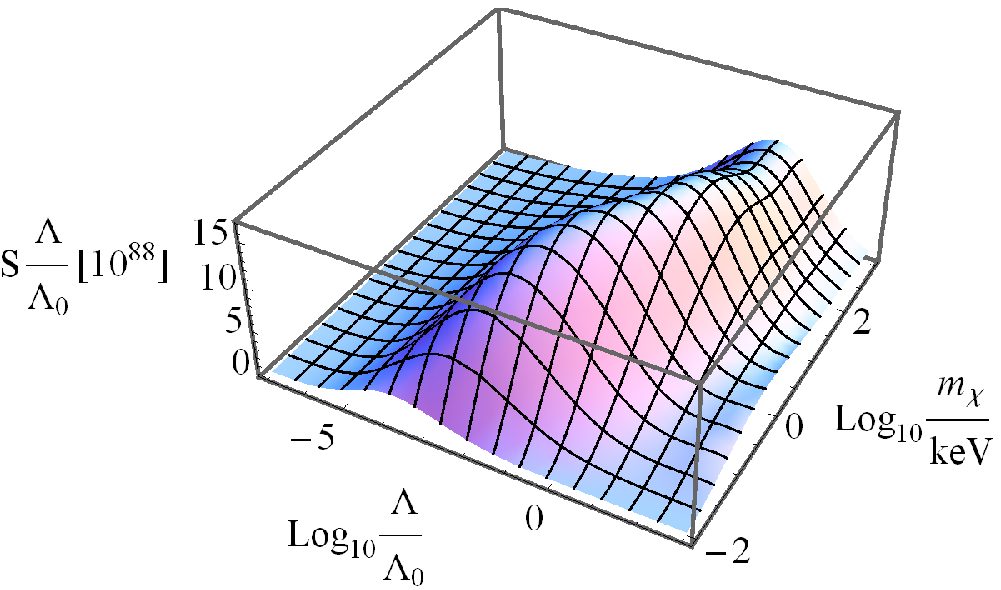}

\caption{Probability of observing a given value of the cosmological constant and mass of dark matter, assuming $g_s = \frac{m_{\chi}}{20 \rm{meV}}$, $\langle \sigma v \rangle = 3 \times 10^{-27} \rm{\frac{cm^3}{s}}$, and that dark matter annihilation is the only
  contributor to entropy production in the causal diamond. This plot assumes the dark matter annihilates into photons which are scattered by
  dust at energies of 20 meV. \label{fig:sigvneg1}}

\end{figure}

\begin{figure}

\includegraphics[width=\linewidth]{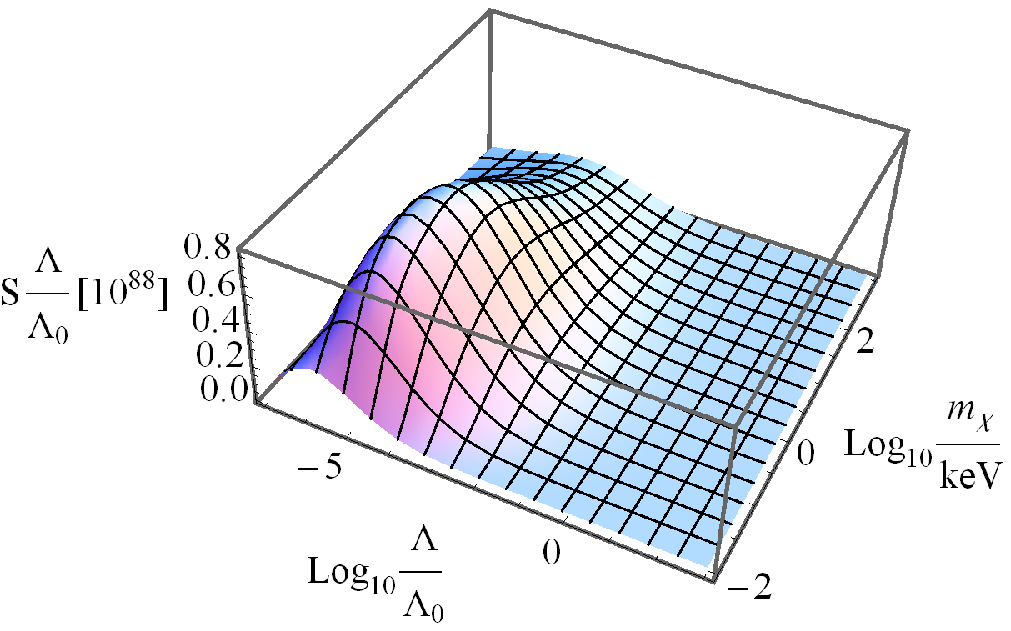}

\caption{Probability of observing a given value of the cosmological constant and mass of dark matter, assuming $g_s = \frac{m_{\chi}}{20 \rm{meV}}$, $\langle \sigma v \rangle = 3 \times 10^{-26} \rm{\frac{cm^3}{s}}$, and that dark matter annihilation is the only
  contributor to entropy production in the causal diamond. This plot assumes the dark matter annihilates into photons which are scattered by
  dust at energies of 20 meV. \label{fig:sigv0}}
	
\end{figure}

\begin{figure}

\includegraphics[width=\linewidth]{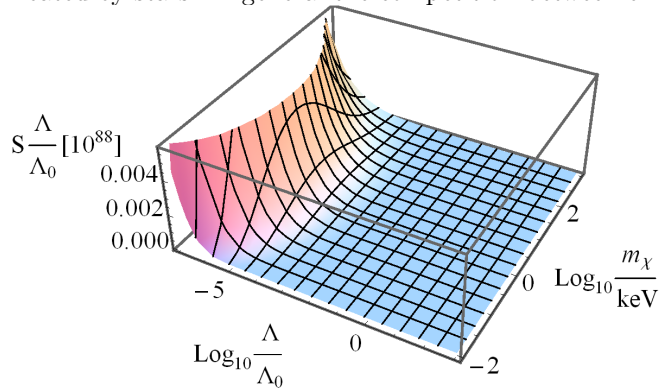}

\caption{Probability of observing a given value of the cosmological constant and mass of dark matter, assuming $g_s = \frac{m_{\chi}}{20 \rm{meV}}$, $\langle \sigma v \rangle = 3 \times 10^{-25} \rm{\frac{cm^3}{s}}$, and that dark matter annihilation is the only
  contributor to entropy production in the causal diamond. This plot assumes the dark matter annihilates into photons which are scattered by
  dust at energies of 20 meV. \label{fig:sigv1}}

\end{figure}

When we vary the cross section, we notice that the effect of increasing the matter abundance is to cause the dark matter to cluster earlier. The probability of observing a universe is maximized when the peak of dark matter annihilation matches the peak of the volume of the causal diamond, so with more matter abundance, the earlier clustering allows us to pick a larger cosmological constant, with roughly the same total entropy production, yielding a runaway preference for huge matter densities as found in~\cite{Maor:2008df}, and shown in 
%Fig.~\ref{fig:varycrossplot} and 
Figs.~\ref{fig:sigvneg2},~\ref{fig:sigvneg1},~\ref{fig:sigv0},~and~\ref{fig:sigv1}. 
%Figs.~\cref{fig:sigvneg2,fig:sigvneg1,fig:sigv0,fig:sigv1}. 
This runaway behavior would suggest that our universe is extremely unlikely; however, as pointed out in\cite{cline2008}, this could be cut off by the formation of black holes which would hide the entropy behind the horizon.

%\begin{figure}
%
%\includegraphics[width=\linewidth]{varycrossplot}
%
%\caption{Contour plot showing contours of the probability of observing a universe as a function of the cross section, mass of dark matter, and cosmological constant. Here we have assumed a constant entropy per annihilation. \label{fig:varycrossplot}}
%
%\end{figure}

%We numerically solve the Boltzmann equation to get the dark matter relic density as a function of the cross section, mass, and degrees of freedom.

%Since this does not match up with our observations, we can conclude that either the CEP is wrong, or that there is additional physics that fixes some of these parameters.

\section{Conclusions}

%{\tt AA: Could you please try and rewrite this without such run-on
  %sentenc(es).  I like the gist of it, but it needs some polishing} 
Our calculations have led to these important points to
take away. First, it appears that light keV scale dark matter is
favored by the causal entropic principle, if we assume that the prior
on entropy per annihilation has only moderate dependence on the mass
of the dark matter particle. (We discuss in Sect. \ref{Sect:Intro} the
relationship to current hints of possible
observed dark matter annihilations.)  Second, since annihilation will peak in 
the far future, a smaller value of the cosmological constant is
favored (when annihilations dominate entropy production), on the order of $10^{-5}$ of the value $\Omega_\Lambda
\approx 0.68$ measured today. Third, dark matter annihilations might
produce a great deal of entropy within the causal diamond, possibly
even greater than the entropy produced from dust heated by stars.  
In general the competition between entropy from dark matter
annihilations and from stars results in a bimodal probability
distribution for $\Lambda$, not a single peak that might be ``tuned''
by these competing effects.  In order for dark
matter annihilation to dominate entropy production we need to have a
large amount of entropy produced per particle decay and a moderately
light dark matter particle.

We also have shown that using the causal entropic principle on the entropy from stars by varying the cross section for dark matter annihilation produces better agreement with observations than varying either the baryon fraction or the total matter abundance independently, as done in~\cite{cline2008}.

Furthermore, we have explored the effect of varying the cross section for dark matter annihilation, and found that the increase in matter density in the early universe resulting from a lowered cross section causes the dark matter to annihilate earlier, allowing a larger value of cosmological constant. This scenario is preferred by the causal entropic principle, and could in principle exhibit runaway behavior, as suggested in~\cite{Maor:2008df}.

\begin{acknowledgments}

We thank M. Brada\v{c} and D. Phillips for helpful conversations. One of us (AA)
would like to thank the 
Kavli Institute for Theoretic Physics at UCSB
for hospitality while this work was 
completed.   This work was supported in part by DOE Grant
DE-FG02-91ER40674 and the National Science Foundation under Grant
No. PHY11-25915. 
\end{acknowledgments}

\bibliography{referencesCEP}

\end{document}